\documentclass{JINST}

\title{PHASES: a concept for a satellite-borne ultra-precise
spectrophotometer}

\author{C. del Burgo$^{a,b}$\thanks{Corresponding
author: Carlos del Burgo (cburgo@uninova.pt)}~, C. Allende Prieto$^c$ and T. Peacocke$^d$\\
\llap{$^a$} UNINOVA-CA3, Campus da Caparica, Quinta da Torre, Monte de Caparica
2825-149, Caparica, Portugal\\
\llap{$^b$} School of Cosmic Physics, Dublin Institute for Advanced Studies,\\
  Dublin 2, Ireland\\
\llap{$^c$}Mullard Space Science Laboratory, University College London, \\
   Holmbury St. Mary, Dorking, Surrey RH5 6NT, United Kingdom\\
\llap{$^d$}Experimental Physics, National University of Ireland, \\
  Maynooth, Co. Kildare, Ireland\\  
}

\abstract{
The {\em Planet Hunting and Asteroseismology Explorer Spectrophotometer},
  PHASES, is a concept for a space-borne instrument to obtain flux
  calibrated spectra and measure micro-magnitude photometric
  variations of nearby stars. The science drivers are 
  the determination of the physical properties of stars and the 
  characterisation of planets orbiting them, to very high precision.
  PHASES, intended to be housed in a micro-satellite,
  consists of a 20 cm aperture modified Baker telescope feeding 
  two detectors: the tracking detector, with a field of 1 degree square, 
  and the science detector for performing spectrophotometry. 
  The optical design has been developed with the primary goal of
  avoiding stray light on the science detector, while providing spectra
  in the wavelength range 370-960 nm with a resolving power that
  ranges from $\sim$900 at 370 nm to $\sim$200 at 960 nm. The signal
  to noise per resolution element obtained for a $V=10$ magnitude star 
  in a 1 minute integration varies between $\sim$ 35 and 140.
  An analysis of the light curve constrains
  the radii of the planets relative to their parent stars' radii,
  which are, in turn, tightly constrained by the combination of 
  absolute spectrophotometry and trigonometric parallaxes.
  The provisional optical design satisfies all the scientific
  requirements, including a $\sim$1\% rms flux calibration strategy 
  based on observations of bright A-type stars and model atmospheres, 
  allowing the determination of stellar angular diameters
  for nearby solar-like stars to 0.5\%. This level of
  accuracy will be propagated to the stellar radii for the nearest
  stars, with highly reliable Hipparcos parallaxes, and more 
  significantly, to the planetary radii.
}


\begin{document}

\section{Introduction}

The finding and characterisation of extrasolar planets is currently one of the
most active research topics. Most discoveries of exoplanets
come from the Doppler-wobble technique, which provides indirect
evidence of the presence of a planet from the analysis of the periodic
variations in the radial-velocity curve of a star, and allows the
determination of $M_p\times\sin(i)$, where $M_p$ is the planetary mass
and $i$ the inclination angle of the orbit. Transiting
exo\-planets are a subset of the exoplanets with $i\sim$90$^\circ$ for
which it is possible to constrain many of their properties. A planet
transit causes a very small but revealing reduction in the apparent
brightness of the host star. This dip in the host star's light curve
(i.e., brightness as a function of time) allows one to determine the stellar
density of the star and the physical parameters of the transiting
planet, such as orbital radius and inclination, and more importantly,
the radius of the planet (see, e.g., Seager \& Mall\'en-Ornelas 2003).
The period of the planet can also be derived when at least two
consecutive transits are observed.

The accuracy in the determination of the properties of the planet from
transit data critically depends on our knowledge of the properties of
the star. For example, for a certain transit depth, the inferred
planetary radius scales in proportion to the assumed stellar radius.
Various methods have been used so far to determine the properties of
 stars with known transits (Torres, Winn \& Holman 2008). For the
stars with precise Hipparcos parallaxes\footnote{The Gaia mission
  will increase by about four orders of magnitude the number of stars with 
  accurate trigonometric parallaxes.} the determination of the stellar 
properties is relatively simple. When parallaxes are not available, 
the measurement and challenging interpretation of the depths and 
shapes of gravity-sensitive absorption features in the star's spectrum 
is usually considered as an alternative. It has been recently shown 
that the uncertainties can be significantly reduced by extracting 
the mean stellar density from the transit light curve 
(see Sozzetti et al. 2007 and Holman et al. 2007).

Transiting planets are much rarer than those found by the Radial Velocity (RV)
technique, due to the statistically low probability of alignment
required to produce a transit signature. The first planet found to
transit was HD 209458b (Henry et al. 2000, Charbonneau et al. 2000),
originally discovered by the RV technique. However, major recent
efforts have made use of a range of wide angle time series surveys to
discover transiting planets from the light curves of millions of stars
(e.g., Horne 2003). Over 60 transiting extra-solar planets are
currently known, most of them discovered by the projects WASP 
(Pollacco et al. 2006), HATNet (Bakos et al. 2004), OGLE (Udalski et a. 2002), 
and recently with the CoRoT mission (Auvergne et al. 2009). 
The most recent transit surveys have improved their detection 
algorithms and follow-up procedures,
and therefore it is expected that transiting planet discoveries will
number in the hundreds over the next few years.

Ground-based photometry is severely limited in precision by the
scintillation noise resulting from air turbulence in the Earth's
atmosphere at altitudes of $\lesssim$4 km. This noise is smaller for
large telescope apertures, but even for the largest 10-m class
ground-based facilities it is extremely difficult to obtain
milli-magnitude (0.1\%) sensitivity in short exposures. Another
problem with ground-based observations is the provision of long-term
(several days or weeks) monitoring of a target, due to the
day-night cycle. Even with polar telescopes or a network of telescopes
sited at different longitudes, the quality of the observations will
nevertheless be highly variable due to changing weather conditions.
Space-borne instruments offer a solution by moving above the
atmosphere, even though the passage through 
the South Atlantic Anomaly (SAA), and solar eclipses by the 
Earth's limb, may reduce the number of 
observations (see, e.g., Sirianni \& Mutchler 2005).
Much current and future investment and planning is
being directed towards space-based facilities designed for, or capable
of, searching for and studying transits, including CoRoT, Kepler, JWST
and ESA's planned Cosmic Vision  mission Plato. These
missions, though crucial, are extremely expensive.  However, known
transits are actually quite bright, and measurements can be made using
far less ambitious instruments. Indeed, to fully study transiting
planet populations in the future, a far more economical means of
obtaining large amounts of space-based data will be preferred. 

Over the last few years, the use of satellites with masses below 100
kg has become feasible, mainly due to improvements in the accuracy
with which the satellites can point  (Walker et al. 2003;
Grillmayer, Falke \& Roeser 2005; Grocott et al. 2006). 
Small, light satellites require much smaller launch vehicles 
than heavy satellites, which considerably
reduces the cost of such missions. It is also possible for  a
microsatellite to ``piggyback'', using spare payload space associated
with larger satellite launches. An early trailblazer for such
micro-satellite technology is the Canadian-led MOST project
(Micro-variability and Oscillations of Stars, Walker et al. 2003),
whose instrument is housed in a suitcase-sized micro-satellite
($\sim$60 kg) and was launched into low-earth sun-synchronous
orbit (820 km altitude, 100 min period) in June 2003.
 For MOST orbit, high levels of cosmic rays due to the passage 
through the SAA and solar eclipses reduce the 
number of observations by $\approx$10 \% 
(Walker et al. 2003, Miller-Ricci et al. 2008).

Ultra-precise space-based photometry provides the critical means to
study known transiting planet systems in a wide variety of ways, both
to better characterise and understand the transiting planets
themselves, and also to search for evidence of additional planets with
sensitivities reaching down to the terrestrial planet range. There
are different areas of research that can be considered, for example
hunting for unseen planets with transit-timing (Agol et al. 2005;
Holman \& Murray 2005), terrestrial planet transits (e.g., Croll et
al. 2007); detailed stellar properties through asteroseismology
(Kjeldsen et al. 2008), planet albedos (Rowe et al. 2006, 2008), 
or moons and rings (Barnes \& O'Brien 2002, Barnes \& Fortney 2004).

 Absolute spectrophotometry can be used to constrain stellar
  angular diameters more accurately than photometric data, and much
  more efficiently than interferometric measurements. An observatory
  in space could provide accurate stellar physical properties, such as
  angular diameters and effective temperatures, for a significant
  sample of bright stars with a modest time investment. These are
  fundamental measurements that are required not only to characterise
  orbiting planets but to derive other stellar properties such as
  ages.  Currently, the only operational spectrophotometer in space
  suitable for accurate measurements of stellar fluxes over a wide
  wavelength range is the Hubble Space Telescope Imaging Spectrograph
  STIS.  The vast costs of operating the Hubble Space Telescope plus
  the high demand for observing time make such a program with STIS
  unfeasible. A simpler instrument on a micro-satellite would
  provide similar or higher accuracy for bright sources at a small
  fraction of the costs.

We present a preliminary optical design for the instrument 
PHASES (Planet Hunting and AsteroSeismology Explorer Spectrophotometer) 
that aims at the determination of the stellar and planetary properties of nearby
transiting systems with unprecedented precision. The telescope feeding
PHASES has an aperture of 20 cm, which will make it possible to obtain
spectra in the wavelength range between 370 and 960 nm with a 
resolving power of  900$>$R$>$200 and a signal-to-noise ratio 
 per resolution element ranging from $\approx$35 to 140 in 
integration times of 1 minute, and achieve a photometric precision 
better than 10 ppm observing two planet transits of stars brighter 
than 10 mag in the Johnson V band 
(assuming a transit length of $\approx$4 hours). 
PHASES will also allow us to tightly constrain the properties of 
the parent stars, with the benefit of a better characterisation of 
the hosted planets.
 The total cost estimated for the entire project 
(design, manufacture, launch and operations) is about 6 million 
Euros, the launch being the most expensive component.

In this paper we outline the conceptual optical design of  the telescope 
and spectrometer PHASES. The description of the microsatellite and
the data handling (compression and download to ground stations) is
beyond the scope of this paper. Here we highlight the crucial
importance of determining the absolute flux level as accurately as
possible to better determine the properties of the star and the
associated planet. In order to achieve that, we have carefully
developed an optical design that offers maximum control of the stray
light. \S \ref{sciencereq} presents the scientific motivation and 
requirements for PHASES. \S \ref{optics} is devoted to 
the preliminary optical design. \S \ref{performance} presents 
the predicted performance of PHASES, and \S \ref{absolutecal} 
describes the flux calibration plan. Finally, \S \ref{summary} 
summarises the paper.

\section{Science drivers and scientific requirements}
\label{sciencereq}

PHASES' main science driver is to dramatically improve our
knowledge of known planetary systems around bright parent stars,
which generally consist of a solar-like star with a close-in giant 
planet in orbit. The main objectives can be summarised:

\begin{itemize}

\item To determine with high precision the physical properties
  (effective temperature, surface gravity, mean density) of the 
  parent stars from the analysis of their spectra and light
  curves (derived from the spectra themselves).  The radii
  of the stars will be constrained from absolute fluxes and parallaxes.

\item To study the seismic activity in the star from the corresponding
  variations in the light-curve.

\item To determine the orbital properties (distance and inclination)
  and physical properties (radius and mass) of the planets with
  unprecedented precision from the corresponding periodic variations in
  the light-curve. A fortuitous discovery of Callisto-like moons 
  might be possible if they eclipse the light from the parent 
  stars at the time of the giant planet transit.

\end{itemize}

These science goals require spectrophotometry from space in order
to improve upon the precision achieved for existing observations. 
Pre-transit and post-transit
spectra will provide information about the properties and level of
activity of the star. The spectral range will be from 370 to 960 
nm, which contains many signatures to characterise the stellar
properties, and also includes activity indicators such as the 
CaII H\&K doublet. The blue part of the spectrum is highly valuable 
to constrain the star's surface gravity, mainly through the Balmer jump.
A minimum resolving power of $\sim$100 is required, but higher
resolution will lead to better results.

The geometry of the transit can be determined from the light curve
during the transit itself, and two consecutive transits make possible
the determination of the period. To observe a transit produced by an
Earth-like planet around a solar-like star requires a photometric
precision of 80 ppm. Our scientific requirement is to achieve 10 ppm
(goal: 3 ppm) for a V=6 mag solar-like star with a transit length of 3
hours. The square root of the transit depth corresponds to the ratio of 
the planetary and stellar radii. A precise determination of the planet
radius depends on a previous accurate determination of the stellar
radius,  to be extracted from the comparison of the observed absolute 
stellar flux and theoretical fluxes at the stellar surface, and 
measured parallaxes. Standard plane-parallel model atmospheres 
can be combined with absolute spectrophotometry to constrain the 
stellar atmospheric parameters for dwarf stars, in order to assign 
an appropriate model atmosphere and theoretical fluxes.

We have carried out simulations of solar-like spectra assuming no 
interstellar extinction, a resolving power of 100 and 
a signal-to-noise ratio per resolution element of about 100.
They show that for the spectral coverage provided by PHASES 
(370 -- 960 nm), the stellar metallicity,
effective temperature, and surface gravity of a solar-like star can be
determined to within 0.2 dex, 50 K, and 0.3 dex, respectively. Using
an independent determination of the metallicity from high-resolution
spectroscopy, it is possible to constrain the effective temperature
and the surface gravity of the star to better than 20 K, and 0.1 dex,
respectively.  An analysis of the solar fluxes compiled by Colina et
al. (1996) using Kurucz (1993) model atmospheres leads to a
solar angular diameter determination of 1928$\pm$40 arcsec (1 $\sigma$), 
only 0.5\% larger than the value from astrometric measurements 
(see Table 2 of Golbasi et al. 2001, or Kuhn et al. 2004). 
Similar agreement is 
obtained in an analysis of the HST-STIS spectrum of Vega
(Bohlin \& Gilliland 2004), giving us confidence that the same
accuracy would be attainable for stars in the A-F-G spectral type range.

 Stellar radii can span a wide range in any given spectral class.
  The effect of metallicity is only secondary --a reduction of about
  0.2 dex from [Fe/H]=0 to -1.5 for a solar-like star-- and the typical
  metallicity span of stars in the Galactic thin disk is about 0.2
  dex.  Much larger variations are found due to evolution as the stars
  leave the zero-age main sequence, especially once they exhaust the
  hydrogen in their cores and begin their ascent on the subgiant
  branch. Consequently the very precise radii measured in eclipsing
  binary systems show scatter in excess of 0.4 dex among unevolved
  (dwarf and subgiant) A and F-type stars with similar surface
  temperatures (see Fig. 2 of Andersen 1991, and also the
  luminosity spread in the Hipparcos field stars discussed by van
  Leeuwen 2009 and references therein). These variations can be
  constrained through spectroscopic determinations of the surface
  gravity, but for a typical accuracy of 0.1 dex in the surface
  gravity, the radius will be uncertain by 0.05 dex, or about 12\%.

The sample of stars to be observed with PHASES would be compiled from
the list published by The Extrasolar Planet Encyclopedia
\footnote{http://exoplanet.eu/}. This list is periodically updated 
and it is expected that will be significantly enlarged in the 
next years. The typical periods of the nearby known exoplanets are 
a few days. We will select a sample of bright stars (V-band magnitude 
below 10 mag) that could be continuously observed during those periods 
(i.e, the stars are inside the Continuous Viewing Zone of the satellite; 
CVZ). 
Given the current list of bright stars, a CVZ with limiting declinations 
of 15$^{\circ}$ and 75$^{\circ}$ is appropriate.  To
  observe the proposed CVZ the microsatellite carrying PHASES should
  be released into a low Earth (850 Km height) Sun-synchronous orbit
  with an inclination to the equator of 133$^{\circ}$.

To determine the transit shape and extract most of the parameters of
the star-planet system, a fast cadence is required, in addition to a
high photometric precision. The determination of an appropriate optical
observing strategy depends upon the detector read time and the
telescope aperture. We discuss this issue further in Sect.
\ref{performance}.

\section{Conceptual optical design}
\label{optics}

PHASES is designed to be carried on a microsatellite launched into
low-earth-orbit. The micro-satellite enclosure  should contain
a cubic science payload of 50-60 cm side, into which the telescope 
and the spectrograph have
to be packed alongside the on-board electronics, computer, and power
supply. Two of the primary mechanical design drivers will be pointing
accuracy so that the science object can be acquired, and pointing
stability to maintain stable illumination of the spectrograph.  There
are two independent detectors,  one CCD for acquisition and
  tracking fed by the telescope and one for science in the
  spectrograph; both are to be passively cooled to just above the 
  ambient temperature of the satellite, estimated to be $\sim220$ K, 
  and stabilised to $\sim230\pm0.1^{\circ}$ K by heating elements on the CCD
  mounting plates (Walker et al. 2003). The science CCD 
  in the spectrograph collects the spectrally dispersed light from
the object of interest.  The function of the tracking CCD is 
to acquire data for the guiding
  and pointing control. The tracking array is read out several times
  during each exposure of the spectrograph CCD and the centroids of
  all sources in the field of view are found. The comparison of the drift
  in the centroids between consecutive readouts is used to maintain
  the position of the science object on the slit via the satellite
  attitude control. In presenting the optical design it is assumed
that the pointing accuracy will be 1$^{\prime\prime}$, which would
meet the requirements. We first present the telescope optics and
imaging on the tracking CCD, then the spectrograph optics.

\subsection{The telescope}

\begin{figure}
  \centering
  \includegraphics[width=7.cm]{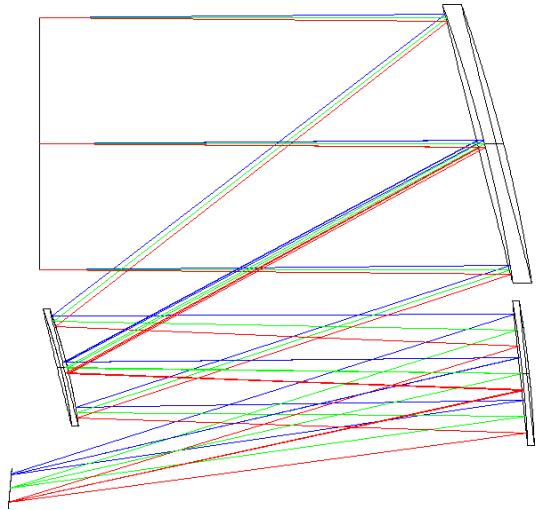}
  \caption{The optical layout of the modified Baker type telescope
    giving unobscured  F5.2 imaging of the $1^\circ$ field of
    view onto the tracking CCD. The three ray bundles focused onto
    the CCD are from sources at the centre and margins of the field of
    view.   The enclosing box measures 30 cm $\times$ 43.6 cm
      $\times$ 23 cm.}
  \label{FigCdB1}%
\end{figure}

Figure \ref{FigCdB1} presents the conceptual telescope layout.  The
design is of a modified Baker type, giving unobscured illumination of
the 1024$\times$1024, 18$\mu$m pixel tracking CCD by the 
$1^\circ$ square field of
view. The aperture stop is 20\,cm in diameter, and it is remote
from the primary mirror. The choice of layout is driven by four
considerations: packaging within the confines of the micro-satellite,
control of stray light on the science detector, maximum light
gathering, and image quality across the field.

A compactly folded, diffraction limited, telescope is chosen to permit
the top of the micro-satellite to be used as an optical bench on which
the telescope optics, tracking CCD and spectrograph are mounted, with
the aperture stop located in the bench. All three mirrors are sized so
that the mirrors for the qualification model, flight spare, flight
model and one spare can be cut from a single parent.  The
  specification of the aperture, field of view and pixel size
  determine the telescope F-number of 5.2, giving a plate scale of
  195$^{\prime\prime}$ mm$^{-1}$ on the tracking detector. The design
is diffraction limited giving an angular diameter for the Airy disc of
 0.92$^{\prime\prime}$ at 370 nm and 2.38$^{\prime\prime}$
at 960 nm.  Distortion is low, reaching $-0.014\%$ at the edge of the
field.

\begin{figure}
  \centering
  \includegraphics[width=9.cm]{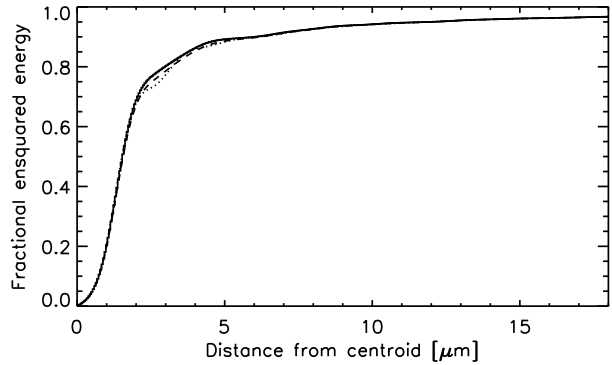}
  \caption{Diffraction ensquared energy: the performance is virtually
    identical across the entire field. Solid and dotted lines
    correspond to the polychromatic and diffraction limit cases,
    respectively.  The image of a point-like source will be $\sim$2/3
      pixels. Displacing the tracking CCD by 100 $\mu$m would spread
      the image of a point source over two pixels.}
  \label{FigCdB2}%
\end{figure}

A graph of the diffraction ensquared energy is shown in Figure
\ref{FigCdB2}. The diffraction performance of the telescope is
essentially constant over the whole field of view. 
 The pixel scale means that the Airy disc of a point-like
  source will be $\sim$2/3 pixels. Defocusing the images on the
  tracking CCD by displacing it 100 $\mu$m would spread the image of a
  point source over two pixels.  Since the purpose of the tracking CCD
  is for centroiding and not imaging this would be an acceptable practise
  giving more accurate centroids.
This will be degraded by the fixed form errors and
mounting tolerances of the three mirrors to give a Strehl ratio of
about $80\%$ in the flight model.

For the telescope, accurate centroiding over the full $1^{\circ}$
field of view is the dominant requirement. Image quality 
 is of critical importance at the spectrograph {\it slit}, where the
smallest possible spread in the light spot is required.  The spectrograph is to be
fed by a pick-off mirror that will collect the science object image
positioned at one corner of the tracking CCD. Positioning the pick-off
at the corner minimises the obscuration of the tracking CCD, whilst
allowing continuous tracking of the science object onto the spectrograph
slit.

\subsection{The spectrograph optics}

\begin{figure}
  \centering
  \includegraphics[width=14.cm]{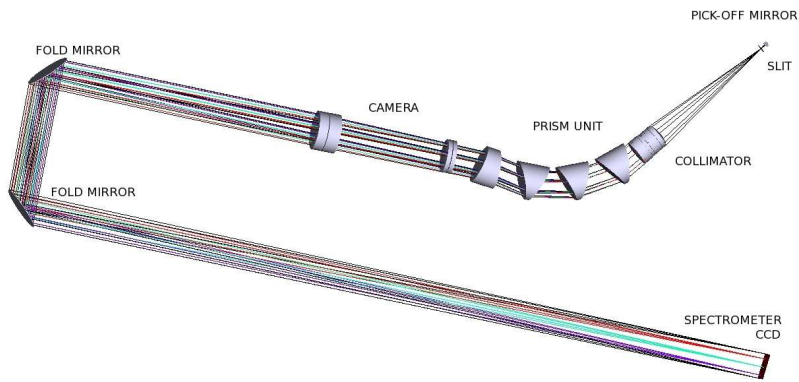}
  \caption{Provisional layout of the spectrograph. The telescope pupil is imaged 
onto the front surface of the prism by the collimating optics (not 
shown) and the spectrally dispersed object is anamorphically imaged onto 
the science CCD by the spectrograph camera, spreading the image  
over ten pixels in the spatial direction.
}
  \label{FigCdB3}%
\end{figure}

The preliminary design of the PHASES spectrograph is shown in Figure
\ref{FigCdB3}. The pick-off mirror positions the point-like science
object on the slit. The optics comprises a collimating lens
  corrected for secondary spectrum, prism unit, and camera; the whole
  being folded to fit into a space envelope measuring 28 cm $\times$ 9
  cm $\times$ 4 cm. The camera is F44, determined by the sampling
  requirement at the spectrograph CCD, the 18 $\mu$m pixel size, and
  the telescope F-number. The tentative width of the slit is
$8.2^{\prime\prime}$ -- the diameter of the fourth minimum of the Airy
pattern at 960 nm -- with an aspect ratio of three. A detailed study
will be carried out from theoretical simulations and laboratory tests
to determine the optimum width and length. The slit has to be sized to
limit variations in the flux reaching the spectrograph detector due to
pointing drifts, while permitting adequate control over stray light,
and avoiding excessive sky background. The slit acts as a field
stop and is the only port through which light may enter the
spectrograph enclosure. For a 8.2$\times$8.2 arcsec$^2$ square slit,
the expected background, dominated by zodiacal light for most
observations, is about $\approx22-23$ mag
arcsec$^{-2}$, or $\approx10^{-4}$ times the flux expected for our
faintest targets (see, e.g., the Cycle 17 HST-STIS Instrument 
Handbook\footnote{http://www.stsci.edu/hst/stis/documents/handbooks/currentIHB/c06\_exptime6.html}). 
There are no geocoronal emission lines in the spectral window 
considered for PHASES.

The obscuration of the tracking CCD by the pick-off mirror will affect
approximately 150$\times$150 pixels of a 1024$\times$1024 pixel array.
A pointing accuracy of $1''$ gives an image displacement of 
  $8.46''$ on the science CCD, corresponding to $\sim2.4$
pixels. Note that this is pointing accuracy, not pointing stability;
at the current time the pointing stability requirement is yet to be
determined and will be a design driver for the satellite control
system. Pointing stability must be maintained over the maximum
  exposure time of 60 seconds. The final pointing stability
  requirement will be determined by the ratio of the F-number of the
  spectrograph camera to the F-number of the telescope and the required
  sampling and image stability at the spectrograph CCD. The
  requirement could be relaxed by reducing the field of view of the
  telescope, thus allowing an increase in the telescope F-number and a
  reduction in the magnification between the focal planes of the
  telescope and spectrograph.

The collimator images the telescope exit pupil onto the second prism
where there will be a stop sized to match the spectrally
  dispersed image of the aperture in the optical bench. 
   Primary control over stray light in the
  spectrograph will be achieved by baffles between the slit and the
  collimator. Additional baffling will be located around the
  spectrally dispersed beam between the camera and the CCD. The
camera is designed to produce an anamorphic image of the star on the
spectrograph CCD with the science object imaged as a ten pixel FWHM
stripe transverse to the direction of spectral dispersion.

\begin{figure}
  \centering
  \includegraphics[width=9.cm]{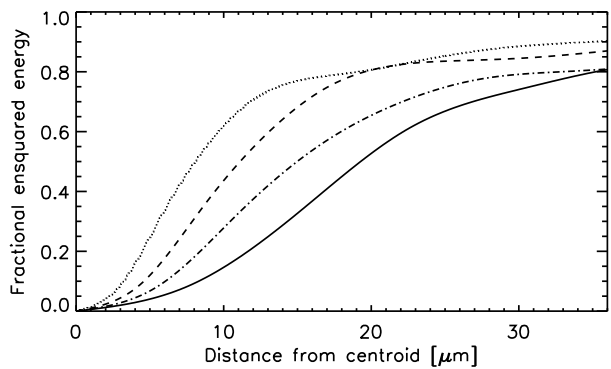}
  \caption{Diffraction ensquared energy of the spectrograph. The
      top lines corresponds to 370 nm, the lower lines to 550, 700 and
      960 nm respectively. The Airy disc diameters are 2 pixels at 370
      nm, 3 pixels at 500 nm, rising to 5.7 pixels at 960 nm.}
  \label{FigCdB4}%
\end{figure}

Although simulations indicate that a resolving power R$\simeq100$ would
suffice to meet the science objectives, a higher resolution will help
to better constrain the atmospheric parameters and is therefore
preferred. The provisional optical design is diffraction limited over
  the full spectral range (see Figure~\ref{FigCdB4}). At the red end of
  the spectrum imaging will be diffraction limited, at the blue end it
  will be pixel limited. The resolution that will be
  obtained in operation will be influenced by the pointing stability,
  wavefront error due to surface form errors in the optics, and
  pixelation. Since the image of the star will be diffraction limited
  in the red and the wavefront errors will be least in the red we
  estimate R$\approx$200 at $\lambda$=960 nm rising to R$\approx$500 at
  $\lambda$=500 nm. At the blue end of the spectrum wavefront
  errors will be greatest and imaging will be pixel limited and we
  estimate R$\approx$900 at $\lambda$370 nm. The actual
  resolution will be determined by laboratory tests.

The system throughput to the spectrograph CCD is estimated to be
  $50\%$ at 370 nm, $62\%$ at 400 nm, and $65\%$ for 500 nm and
  above.  The final transmission will be greatly influenced by the
choice of coatings used, giving some freedom to bias the throughput
towards one end of the spectrum. With a dielectric coating on the
mirrors, the throughput of the telescope could be raised above $95\%$,
and biasing the efficiency of the coatings on the spectrograph optics
towards the blue can be used to partially offset the lower quantum
efficiency of the detector. The final choice of coatings and glasses
will be made later in the design cycle.

\subsection{Data acquisition}\label{acquisition}

The dispersed image of a star on the science detector or aperture will
be sampled by 1024 pixels in the dispersion axis and 10 pixels
in the perpendicular axis. The latter number may be modified
according with current simulation and laboratory tests. 
The mean dispersion is 0.58 nm pixel$^{-1}$. The expected resolving 
power changes from $\sim$900 at 370 nm to $\sim$200 at 960 nm, which
results in a higher resolution towards blue wavelengths, in line
with the science needs.

The maximum exposure time is of 60 seconds, which allows us
to obtain a minimum signal-to-noise per pixel of about 25 
(i.e., 35 per resolution element at 370 nm) 
for the faintest (V=10) stars to be observed (see \S \ref{performance}).
The pixel binning strategy may be changed to optimize the observation
of faint objects at lower resolving powers. 
A binning in the dispersion axis can be used 
to preserve the read noise per bin, increasing the signal-to-noise
ratio.

The aperture (i.e. image of the spectrum on the CCD) should be
defined, traced along the dispersion axis and extracted. It is
required that the aperture of the same object may move only slightly
during observations to reduce the impact of the pixel-to-pixel
sensitivity variations in the detector on the determination of the small
variations that are associated to the transit of a planet or the
activity of a star.

A data reduction pipeline would be developed for PHASES, where
problems such as the correction of cosmic rays, pointing instability
and stray light subtraction will be sorted out 
(see, e.g., Huber \& Reegen 2008). 
This would be discussed in a forthcoming publication
once the simulations and laboratory tests of PHASES are completed.
We highlight here that the use of spectrophotometry helps
to clean up the images from cosmic rays, bad columns and dead pixels,
since it is possible to make a comparison of the observed spectra
with the state-of-the-art synthetic models, which mimic closely
the observations (see Sect. \ref{abscalsect}). This is a significant 
advantage for performing ultra-precise measurements.

\section{Instrument performance}\label{performance}

\begin{figure}
  \centering
  \includegraphics[width=9.cm]{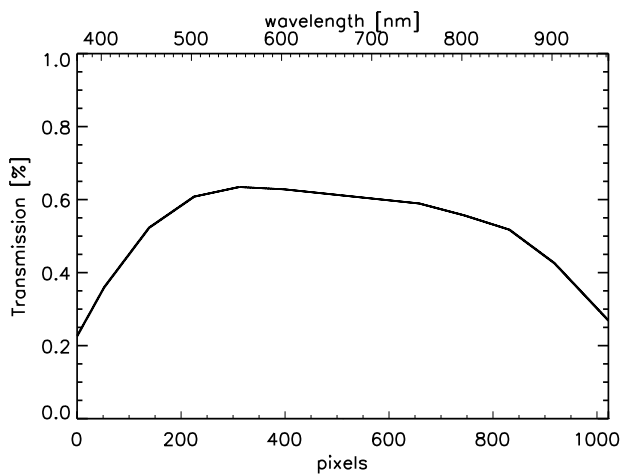}
  \caption{Predicted total transmission of PHASES versus wavelength.}
  \label{FigCdB5}%
\end{figure}

\begin{figure}
  \centering
  \includegraphics[width=9.cm]{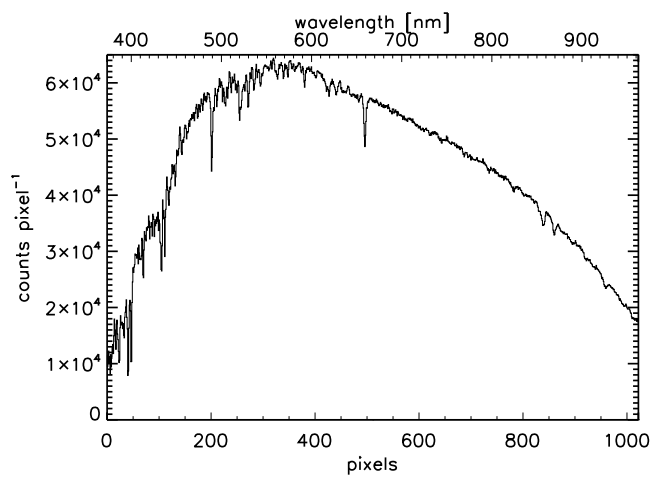}
  \caption{Spectral energy distribution prediction (in counts per photon) 
for the star HD 209458 as seen by PHASES with an exposure time of 60 seconds. 
Poisson noise has been added. Note the change in resolution from the blue 
to the red part of the spectrum}.   
  \label{FigCdB6}%
\end{figure}

The performance of PHASES has been estimated assuming that the
telescope mirror transmission and the spectrograph
specifications are as given in Sect. \ref{optics}.
The assumed quantum efficiency of the science detector is
$\approx$40\% at 370 and 960 nm, with a maximum approaching
$\sim$97\% at 550 nm, and values higher than 90\% in the wavelength
range between 480 and 750 nm. These capabilities can be
provided by commercial back-illuminated deep depletion silicon
CCDs (see, e.g., http://www.e2v.com). 
Fig. \ref{FigCdB5} shows the total predicted throughput 
as a function of wavelength and position on the chip.

The scientific requirement is to achieve a photometric
precision of 10 ppm (goal: 3 ppm) using the full
spectral range of PHASES for a V=6 mag solar-like star with a transit
length of 3 hours. Our simulations show that it is possible to reach
a photon-noise-limited precision of 2 ppm given the expected
performance of PHASES. The only source of noise considered here is
shot noise that is given by the square-root of the number of photons
(assuming a gain of 1 count per photo-electron) and follows a Poisson
distribution. The expected dark signal ($\leq$0.01 e$^-$ s$^{-1}$
pixel$^{-1}$) and read noise ($\leq$5 e$^-$) of the detector will
be negligible for the full sample of interest in the typical
exposure times. 

The quality of acquired data will be degraded by satellite pointing
jitter, pixel-to-pixel variations, dead or hot pixels and bad columns.
These defects will adversely influence our ability to isolate variations
intrinsic to the star (e.g., stellar activity) or due to planetary transits.
The study of the influence of these effects, and their mitigation,
will be the subject of a future report.

HD 209458 (G0V, V=7.65) is the brightest star with a known planet in
orbit. For this object, a photon-noise-limited photometric precision
of 5.7 ppm would be achievable in the 3.7 hour duration of the transit
of planet HD 209458b. With two transits it would be possible to achieve
a precision of 4.0 ppm. Fig. \ref{FigCdB6} shows a simulation of
an observed spectrum (once the aperture has been extracted and
spectrally calibrated) of the star HD 209458. 
Note how the resolution decreases from the blue to the red 
wavelengths. The minimum signal-to-noise per pixel for 
a single 1-minute spectrum of HD 209458 varies 
from $\sim$70 at 370 nm to $\sim$225 at about 960 nm. The signal-to-noise
per resolution element is $\approx$100 at 370 nm and $\approx$380 at 960 nm.
The star XO-3 with a V-band magnitude 
of 9.8 and a planet transit (Johns-Krull et al. 2007) that lasts 
3.7 hours is as bright as the faintest stars intended to be observed, 
with an expected photon-noise-limited photometric precision of 
10.9 ppm after two transits. A single 1 minute spectrum
of XO-3 has a signal-to-noise per pixel that varies from $\sim$25 at 370 nm
to $\sim$85 at 960 nm, with a maximum of 80 at 500 nm. The signal-to-noise
per resolution element is $\sim$35 at 370 nm, 
130 at 500 nm and 140 at 960 nm.

\section{Absolute calibration}\label{absolutecal}
\label{abscalsect}

The accuracy of the inferred planetary radius will depend upon the
accuracy of the stellar radius; it is necessary to precisely
characterise the parent star in order to characterise its planets.
Stellar angular diameters can be determined from interferometry or by
comparing predicted fluxes at the stellar atmosphere with those
observed from Earth (provided an accurate flux calibration). With
accurate angular diameters and distances (trigonometric parallaxes
from Hipparcos/Gaia), we can determine accurately the properties of
the parent stars.

\begin{figure}
  \centering
  \includegraphics[width=9.cm]{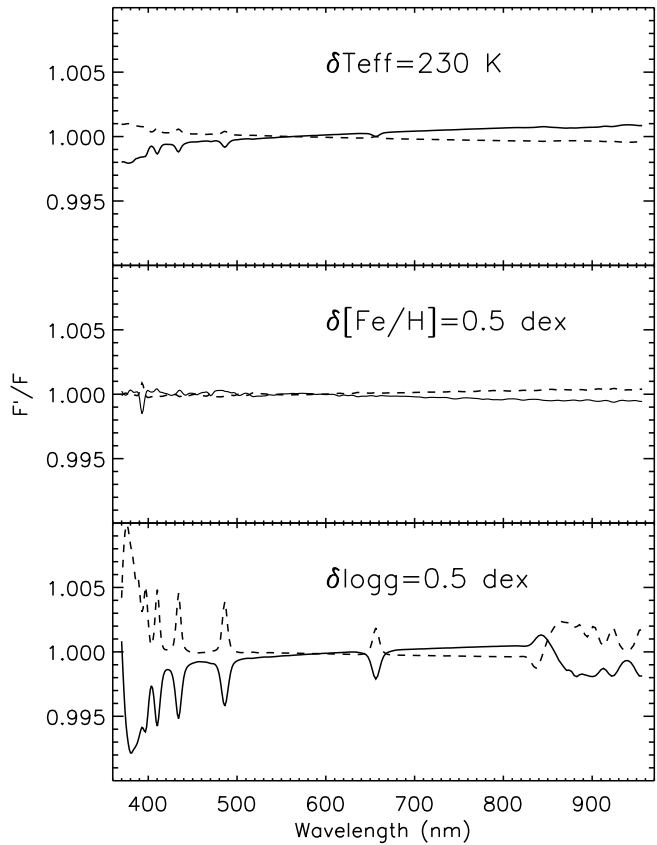}
  \caption{Relative variations in the stellar flux emerging from a
    Vega-like star associated with changes in the atmospheric
    parameters of $\delta T_{\rm eff}= \pm$230 K (upper panel),
    $\delta$[Fe/H]$= \pm 0.5$ dex (middle panel), and $\delta\log g=
    \pm 0.5$ dex (bottom panel). The solid lines correspond to the
    ratio between the flux after a reduction in one of the parameters
    and the original flux, while the dashed lines correspond to positive
    increments in one of the parameters.}
  \label{FigCdB7}%
\end{figure}

\begin{figure}
  \centering
  \includegraphics[width=9.cm]{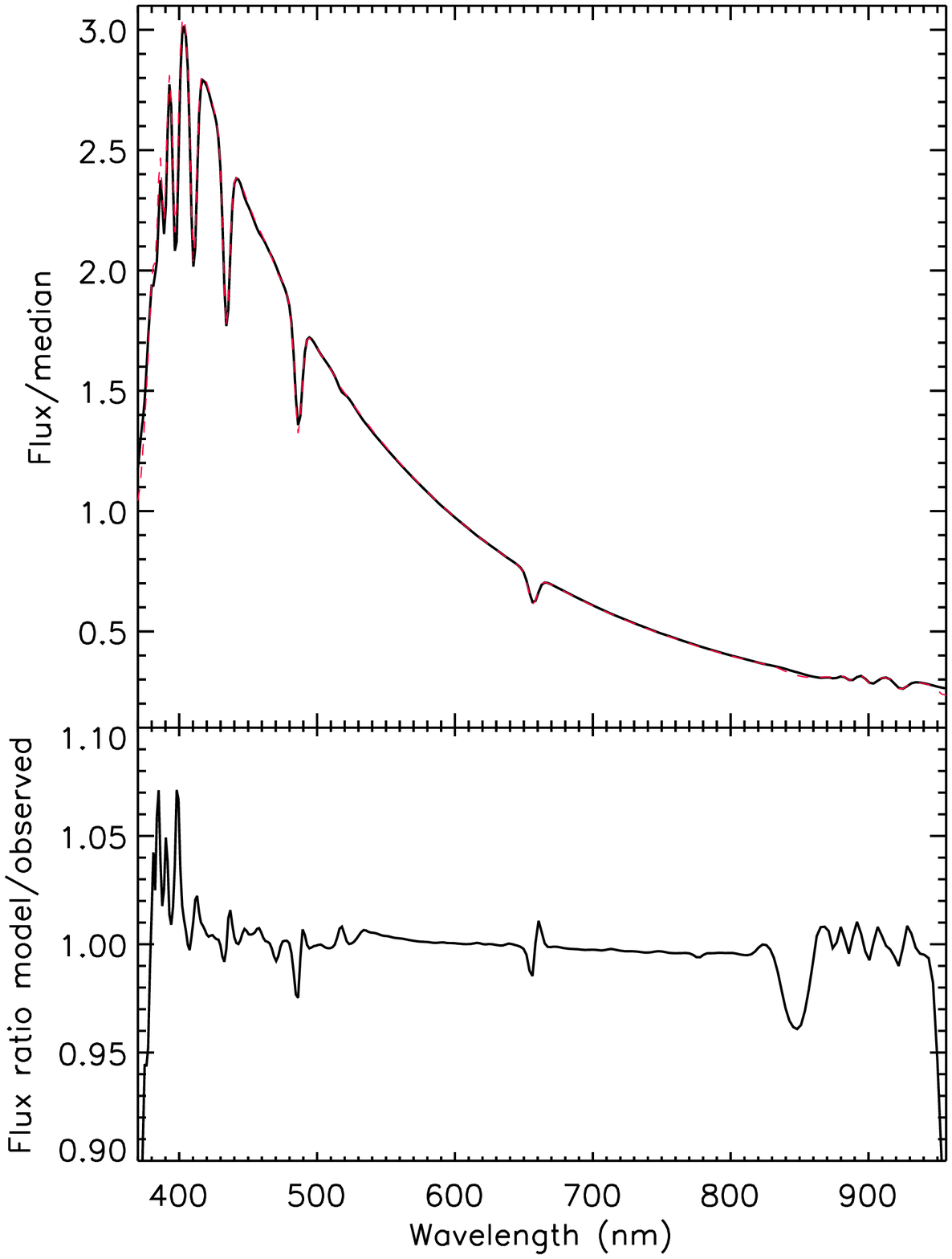}
  \caption{Vega fluxes in the PHASES spectral window normalized by
    their median value. All lines are for a resolving power of
    $R=100$. The black solid and red broken lines in the upper plot show the
    observed spectrum (HST/STIS; Bohlin \& Gilliland 2004) and the 
    best-fitting model, respectively. The lower panel shows the ratio of the two.}
  \label{FigCdB8}%
\end{figure}

There are two distinct methodologies for performing
spectrophotometric calibration of astronomical observations. 
Most catalogues of spectrophotometric standards are
derived from ground-based observations, carefully tied to absolute
fluxes from a controlled source (see, e.g., Hayes 1985).

Alternatively, the spectrophotometric catalogue of the HST (FOS, STIS
and NICMOS; see Bohlin 2007 and references therein) establishes the
relative spectral energy distribution for a number of standards using
a different strategy, based on theoretical, model-atmosphere,
predictions. The absolute zero point of the HST scale is ultimately
tied to the Vega V-band magnitude (see Colina \& Bohlin 1994, 
Bohlin \& Gilliland 2004, and 
references therein). The HST primary
standards, for which the spectral shape is postulated based on model
atmospheres, are three DA white dwarfs. Their atmospheric parameters
(surface gravity and temperature) are defined from the analysis of the
shape of the Hydrogen lines, which are highly sensitive to both the
electron density (related to temperature) and the gas pressure (which
correlates with the surface gravity). HST is free of atmospheric
extinction, and the degradation of the charge transfer efficiency 
in its CCD's has been studied in detail (Goudfrooij et al. 2006).

This second approach relies on the premise that the atmospheres of
these objects, made of pure hydrogen, are simple enough to model with
high accuracy. With three white dwarf stars (G 191 B2B, GD 153 and GD 71),
Bohlin reports an internal consistency at a level of $<$0.5\% for STIS
(near-UV and optical), and about 1\% for NICMOS observations (IR). The
STIS spectrum of Vega (Bohlin \& Gilliland 2004), calibrated upon the
DA white dwarfs, closes the loop. A model atmosphere for this star 
matches the observations to 1--2 \%, and a satisfactory agreement 
is also found with the absolute fluxes of Hayes (Bohlin 2007).

The HST primary standards are too faint for a small aperture telescope
like the one coupled to PHASES; 
The HST DAs are among the brightest white dwarfs known. Other DA could
be analysed spectroscopically and used as standards, but there are
none brighter than $V<9.5$, and therefore they are unsuitable for
PHASES.

The success in matching Vega's spectrum with models found by Bohlin,
suggests that A-type stars could also be used as standards. 
A-type stars have
their continuum shaped by bound-free (photoionization) of neutral
hydrogen atoms. Their atmospheric opacity is significantly affected 
by metal lines in the UV, but that is not the case for the optical 
and near-IR. In addition, at high-enough temperatures the vast majority 
of the electrons in these atmospheres are coming from the ionization of
hydrogen. Furthermore, with lower densities than in white dwarfs,
convection (an important source of uncertainty in the models) is
inhibited at lower temperatures.

The published values for the temperature of
Vega show a maximum range of about 230\,K (see Garc\'{\i}a-Gil et al.
2006), and the surface gravity of any bright A-type star with trigonometric
parallaxes from Hipparcos should be known to better than 0.2 dex. 
Fig. \ref{FigCdB7} illustrates that such a range in $T_{\rm eff}$, as
well as very large variations of 0.5 dex in
[Fe/H] or $\log g$, have only a very limited effect on the
predicted fluxes over the spectral range of PHASES.  
The effect of metallicity, which should be possible to
determine from existing (or easily obtained) ground-based
high-resolution spectra to within $\sim$0.1 dex, is, in fact, 
completely negligible. Consequently, bright A-type stars will be 
suitable for flux calibration of PHASES.

Given the extremely low sensitivity of the continuum to metallicity,
we have attempted to determine T$_{\rm eff}$ and logg (fixing the metallicity)
from the STIS spectrum of Vega smoothed to R=100, similar to the experiments
described in \S \ref{sciencereq} for the solar spectrum. 
The result is illustrated in Fig. \ref{FigCdB8}, 
and the agreement is excellent with an RMS scatter of just 
0.6\% between 400 and 800 nm.
We note that Vega is known to be a fast rotator (see, e.g., Aufdenberg
et al. 2006), although this has a very limited impact in the PHASES
wavelength range.

With an RMS scatter of 0.012 (1.2\%), which we find when matching the 
solar spectrum with calculations based on Kurucz models 
(see \S \ref{sciencereq}), 
and 2--3 pixels per resolution
element, the scaling factor between the computed and observed fluxes
can be determined to $\sim$ 0.1 \%. Thus, the limiting factor for
absolute calibration will be the zero point. If the calibration is
anchored to Vega's fluxes, this is given by the uncertainty in its V =
0.026 $\pm$ 0.008 mag (Bohlin 2007). Our best fitting model
([Fe/H]$=-0.7$, imposed; $T_{\rm eff}$=9506$\pm$23\,K; $\log g$=
4.01$\pm$0.02 dex) predicts a flux of $5.40 \times 10^7$ erg s$^{-1}$
cm$^{-2}$ \AA$^{-1}$ at 555 nm, or, scaled by the inferred angular
diameter ($\theta=3.31$ mas), $3.48\times 10^{-9}$ erg s$^{-1}$
cm$^{-2}$ \AA$^{-1}$ at 555 nm at Earth. This result is in
excellent agreement with the figures from the STIS observations
provided by Bohlin of $3.46 \times 10^{-9}$ erg s$^{-1}$ cm$^{-2}$
\AA$^{-1}$.

Provided the uncertainties in the atmospheric parameters for the
calibration stars are comparable to those we inferred for the Sun or
Vega, they will have a minor impact on the predicted fluxes. If the
accuracy in the flux calibration is translated to the PHASES targets
undisturbed by frequent observations of the standards, then such
observations have the potential to lead to measurements of the angular
diameter of the targets to a $\sim 1$ \% level. This figure compares
favourably to (model-based) photometric calibrations (Ram\'{\i}rez \& Mel\'endez
2005; Casagrande 2008), or even interferometric measurements for most
dwarfs (see, e.g., Baines et al. 2009). For bright stars with accurate
parallaxes, we can expect the uncertainty in the estimated angular
diameter to be retained in the implied radii to be used to derive the
planets radii.

\section{Summary}
\label{summary}

In this paper, we have outlined the motivation and concept for
the PHASES instrument, which is designed to perform ultra-precise 
relative photometry and absolute spectrophotometry from a small-sized 
space platform. PHASES is optimised to provide a detailed characterisation 
of the transiting planets in orbit around nearby stars.
 
The proposed instrument has a dual objective: to measure the light
curves of the target systems (stars brighter than V=10) during 
the transits with a precision better than $\sim 10^{-5}$ (10 ppm), 
and to obtain spectrophotometry of the parent star
with an absolute accuracy of $\sim1$\%. These two measurements, together 
with absolute parallaxes available from Hipparcos and later Gaia, can 
place tight constraints on the orbital elements and the radius of the 
transiting planet. The goals can be realised by an opto-mechanical 
design optimised to control scattered light, by dispersing the target's 
light over a large area of the detector, and minimising the 
cross-contamination of the spectrum by nearby sources. The avoidance of 
spectral cross-contamination is a critical driver for both the optical 
and the mechanical designs. This consideration leads to the adoption of 
an unobscured quasi-Baker telescope design, and the use of the slit as a 
spacial filter separating the acquisition and guiding function of the 
telescope optics from the spectrograph optics. Within the spectrograph 
enclosure, careful attention can be given to stray light control.

We argue that bright A-type stars can be used for flux calibration in
the same way that DA white-dwarfs are used for fainter sources in many
space-based instruments; by using model atmospheres and scaling
the fluxes using broad-band photometry.

For solar-like stars, by determining the metallicity of the parent
star using high-resolution spectroscopy, it is then possible to use
$R\geq 100$ spectrophotometry to tightly constrain the effective
temperature and the surface gravity. Assuming model atmospheres
 match the spectral energy distribution of real stars to better than 1 \%, 
having absolute fluxes accurate to $\sim1$ \% will directly 
provide $\sim 0.5$\% angular diameters and, for nearby stars, 
a similar accuracy in the stellar radius. It has been
demonstrated that space-based light curves can be used to determine
the stellar density and the ratio of the planetary to the stellar ratio 
with high precision (see, e.g. Seager \& Mall\'en-Ornellas 2003,  
Mandel \& Agol 2002), and therefore the availability, from a single 
instrument, of the two measurements, will lead to a significant improvement 
in the final accuracy of the inferred planetary radii, and the properties
of the planetary systems in general.

\acknowledgments

The authors would like to thank David Pinfield and Dave
Walton for useful discussions, and an anonymous referee 
for thoughtful comments.

\end{document}